\documentstyle[12pt]{article}

\def\beq{\begin{equation}}
\def\eeq{\end{equation}}
\def\bea{\begin{eqnarray}}
\def\eea{\end{eqnarray}}

\begin{document}

%%%%%%%%%%%%%%%%%%%%%%%%%%%%%%%%%%%%%%%%%% TITLE, AUTHORS and ABSTRACT

\title{Effective Lagrangians for physical degrees of freedom in
the Randall-Sundrum model}

\author{Edward E. Boos$^{a}$, Yuri A. Kubyshin$^{a,b}$,
\and
Mikhail N. Smolyakov$^{c}$ and Igor P. Volobuev$^{a}$ \\
\  \\
$^{a}$Institute of Nuclear Physics, Moscow State University \\
119899 Moscow, Russia \\
$^{b}$Departament MA IV, Universitat Polit\`ecnica
de Catalunya \\
M\`od. C-3, Campus Nord, Jordi Girona, 1-3 \\
08034 Barcelona, Spain \\
$^c$Physics Department,  Moscow State University \\
119899 Moscow, Russia}

\maketitle

\begin{abstract}
We derive the second variation Lagrangian of the
Ran\-dall - Sun\-drum model with two branes, study its gauge invariance
and diagonalize it in the unitary gauge.
We also show that the effective four-dimensional
theory looks different on different branes and calculate
the observable mass spectra and the couplings of the physical
degrees of freedom of 5-dimensional gravity to matter.
\end{abstract}

%%%%%%%%%%%%%%%%%%%%%%%%%%%%%%%%%%%%%%%%%%%%%%%%%%% BODY of the article

\section{Introduction}

The Randall-Sundrum (RS) model with two branes, first proposed and
analyzed in Refs. \cite{RS1,RS2}, has been extensively  studied  in the
literature, see Refs. \cite{Ant01} - \cite{Kub01} for reviews.
The interest in this model, known as the RS1 model, is due to a novel
solution of the hierarchy problem suggested within it, and to a
possibility to check its predictions in the planned collider experiments
(see articles \cite{DHR1,DHR2} and Refs. \cite{Kub01,TDR,Bes} for reviews).

The RS model describes gravity propagating in a five-dimensional
space-time
with the fifth dimension being compactified to the orbifold $S^{1}/Z_{2}$
with the circumference of the circle equal to $2R$. Let us denote
coordinates of the space-time by $\{\hat{x}^{M}\} = \{ (x^{\mu},y)\}$,
$M=0,1,2,3,4$, so that $\{x^{\mu}\}$, $\mu = 0,1,2,3$ are the coordinates of
the four-dimensional space-time and $\hat{x}^{4} \equiv y$ is the
coordinate along the fifth dimension.
$Z_{2}$-symmetry of the orbifold is realized by identifying points
$(x^{\mu},y)$ and $(x^{\mu},-y)$, where $y$ is a point of the circle.
In addition we have the usual periodicity condition which identifies
$(x^{\mu},y)$ and $(x^{\mu},y+2R)$. The branes are located at the fixed
points $y=0$ and $y=R$ of the orbifold.

The action of the RS1 model is equal to
\[
S = S_{g} + S_{1} + S_{2},
\]
where $S_{g}$ is the five-dimensional Einstein action with the
cosmological term,
\beq
S_g = \frac{1}{16 \pi \hat G}
\int \left( {\cal R} - \Lambda \right) \sqrt{-g} d^{5}\hat{x}, \label{RSg}
\eeq
and $S_{1}$ and $S_{2}$ are the standard actions of the branes:
\bea
 S_1&=& V_1 \int \sqrt{- \tilde g} \delta(y) d^5 \hat{x},  \label{RSb1} \\
 S_2&=& V_2 \int \sqrt{- \tilde g}  \delta(y-R) d^5 \hat{x}.
\label{RSb2}
\eea
Here $\hat{G}$ is the five-dimensional gravitational constant,
$g_{MN}$ is the metric tensor in five dimensions
with the signature $(-,+,+,+,+)$, and  $\tilde g=\det (g_{\mu\nu})$ is
the determinant of the induced metric.

The exact background solution for the metric in
this model is of non-factorizable type and is given by \cite{RS1}
\beq
ds^{2} = \gamma_{MN} d\hat{x}^M d\hat{x}^N = e^{2\sigma(y)} \eta_{\mu\nu}
{dx^\mu dx^\nu} + (dy)^{2},
   \label{metricrs}
\eeq
where $\eta_{\mu \nu}$ is the Minkowski metric.
The function $\sigma (y)$ is equal to $\sigma (y) = -k|y|$ for
$-R \leq y \leq R$ and is continued periodically outside this interval.
The parameter $k$ has the dimension of mass, and the cosmological
constant $\Lambda$ and the brane tensions $V_{1}$, $V_{2}$ are
related to it as follows:
\[
\Lambda = -12 k^2, \quad V_1 = -V_2= -\frac{3k}{4\pi \hat G}.
\]
If $k > 0$ the brane at $y=0$ (brane 1) has positive tension and
the one at $y=R$ (brane 2) has negative tension.

Though many studies of this model have been done already, some issues have
not been properly understood and clarified. Thus earlier
papers dealt only with the equations of motion for the linearized gravity,
and its Lagrangian description remained practically untouched.
In the present paper we first repeat the derivation of the Lagrangian of
quadratic fluctuations around the background solution (\ref{metricrs}) and
carry out its diagonalization. We also discuss the problem  of
imposing a gauge condition in a consistent way and introduce the unitary
gauge, in which the radion is represented explicitly by a
four-dimensional scalar field.

In the majority of papers on the RS1 model masses and couplings
constants in effective theories on the branes are calculated
with respect to the induced metric. However, as one can see
from Eq. (\ref{metricrs}), the induced metric on brane 2 contains
the conformal factor $e^{2\sigma (R)} = e^{-2kR}$. Hence the
coordinate system $\{x^{\mu}\}$ is not Galiliean on the brane
(coordinates are called Galillean if $g_{\mu \nu}=diag (-1,1,1,1)$,
see for example Ref. \cite{LL}) and, therefore, the naive
identification of parameters from the Lagrangian leads to
wrong conclusions. This issue was recognized and discussed in Ref.
\cite{GNS}. In this  paper, a covariant procedure was proposed,
which consists in using
covariant equations and expressing all coordinate distances
in terms of proper distances. For example, within
this approach the masses of brane fields, i.e. fields localized
on a brane and not propagating in the five-dimensional bulk, were
determined from the rate of exponential falloff of the 2-point
function. It was shown that such masses measured by brane observers
are independent of the warp factor and are given by the Lagrangian
masses. In Ref. \cite{GNS} it was also confirmed that Newton's
constant of gravitational interaction between two masses depends
on the warp factor in the same way as in the conventional interpretation
\cite{RSgrav}. Also, masses of bulk scalar fields were analyzed there.
It was found that an observer on the negative tension brane sees
particles with masses $m_{n} \sim k$, whereas an observer on the
positive tension brane detects exponentially suppressed masses
$m_{n} \sim k e^{-kR}$.

In the present paper we re-examine these issues and calculate the
physical masses and coupling constants on each of the branes in
a more straightforward and, as we believe, simpler way. Instead
of analyzing the 2-point function and the geodesic equation we
pass to the Galilean coordinates and perform necessary
rescalings of fields. This allows us to get the correct identification
of the physical parameters in the effective theory on the branes. We
calculate the mass spectra and couplings
of the Kaluza-Klein modes of the gravitons to matter
on the branes, as well as the masses of brane fields.

As it is usual in scenarios considered in the literature,
we assume that our brane is brane 2 (the one with negative tension)
and our considerations lead to a phenomenology very similar to the one
discussed in Refs. \cite{DHR1,DHR2}.
However, the general physical picture of
the brane world appears to be quite different. Thus, in our approach
the fundamental physical scale, i.e. the five-dimensional "Plank mass"
$M$, should be in the TeV range. It is very natural to have the  value for
the parameter $k$ in the same TeV range. The four-dimensional
Plank scale $M_{Pl}$ is the effective scale for the gravitational zero mode
interaction on brane 2. The hierarchy is generated by the exponenential
term coming from the warp factor $e^{2\sigma(y)}$ in Eq. (\ref{metricrs}).
Namely, the Planck mass is related to the fundamental scale
by the relation
\[
  M_{Pl}^{2} \approx \frac{M^{3}}{k} e^{2kR},
\]
see Refs. \cite{Rub01,BKSV} and Sect. 4 of the present article.
To reproduce the value of $M_{Pl}$ the size $R$ has to be
chosen such that $kR \approx 30 \div 35$.
Therefore, the size of the extra dimension is expected to be about
35/TeV\footnote{As an alternative interpretation this possibility
was mentioned in Ref. \cite{RS1}.}, much larger than 35/$M_{Pl}$
discussed in earlier papers. We will show that
for these values of the model parameters the coupling of
gravity to matter on the "hidden" brane (brane 1) turns out to be very
strong, namely the gravitational constant is proportional to $M^{-1}$ and
not to $M_{Pl}^{-1}$, as on brane 2. There are also some differences in the
radion interactions which will be discussed below.

The article is organized as follows. In Sect. 2 we review
gauge transformations in an arbitrary background and derive
an expression for the Lagrangian of quadratic fluctuations
which will be convenient for our purposes. In Sect.~3 we use
this result for the calculation of the Lagrangian of quadratic
fluctuations in the RS1 model, repeat its diagonalization in
a systematic way paying attention to the gauge transformations
and identify physical degrees of freedom. In parallel we
prove the decoupling of the classical equations of motion for
these degrees of freedom. In Sect.~4 we study both
brane scalar fields and the bulk fields
arising from a five-dimensional metric. We calculate their
masses and couplings in effective theories on the branes using
the Galilean coordinates. Sect.~5
contains some discussion of the results and concluding remarks.

\section{Second variation Lagrangian in a general background}

For the purpose of generality let us consider the standard gravitational
action with the cosmological
constant in $(4+d)$-dimensional space-time with coordinates $\{\hat{x}^M\},
M=0,1, \cdots d+3$:
\begin{equation}
S_g = \frac{1}{16 \pi \hat G} \int \left({\cal R}-\Lambda\right)
\sqrt{-g}\, d^{4+d}\hat{x},   \label{actiong}
\end{equation}
where $\hat{G}$ is the multidimensional gravitational constant,
${\cal R}$ is the scalar curvature and $g_{MN}$ is the metric
with signature $(-1,1, \cdots 1)$. For $d=1$ this expression coincides
with five-dimensional action (\ref{RSg}).

Let $\gamma_{MN}$ be a fixed background metric. We denote
$\hat \kappa = \sqrt{16 \pi \hat G}$ and parameterize the metric
$g_{MN}$ as
\begin{equation}
  g_{MN}(\hat{x})= \gamma_{MN}(\hat{x}) + \hat \kappa h_{MN}(\hat{x}).
  \label{metricpar}
\end{equation}
If we substitute this formula into (\ref{actiong}) and retain
only the terms of the zeroth order in $\hat \kappa$, we
get the following Lagrangian, which is  usually called the
second variation Lagrangian:
\begin{eqnarray}
L^{(2)}_g/\sqrt{-\gamma} & =& -\frac{1}{4}\left(\nabla_R h_{MN}
\nabla^R h^{MN} - \nabla_R h \nabla^R h + 2\nabla_M h^{MN}\nabla_N
h -\right.   \nonumber \\ &-& \left. 2 \nabla^R h^{MN} \nabla_M
h_{RN}\right) + \frac{1}{4}\left({\cal R}-\Lambda
\right)\left(h_{MN} h^{MN} - \frac{1}{2} hh\right) +   \nonumber
\\ &+&  {\cal G}^{MN} h_{MR} h_N^R - \frac{1}{2}{\cal G}^{MN} h_{MN} h,
\label{Lagrangian0}
\end{eqnarray}
where $\gamma = \mbox{det} \gamma_{MN} $, $h = h_M^M$, and
${\cal G}_{MN}= {\cal R}_{MN} - \frac{1}{2} \gamma_{MN}
\left({\cal R}-\Lambda\right)$. The Ricci tensor
${\cal R}_{MN}$, the scalar curvature ${\cal R}$ and the covariant
derivative $\nabla_M$ are calculated with respect
to the background metric $\gamma_{MN}$.
This formula in rather complicated notations can be found
in \cite{BDW}.

Action (\ref{actiong}) is invariant under general  coordinate
transformations $\hat{y}^M = \hat{y}^M(\hat{x})$,
the corresponding transformation of the metric being
\begin{equation}\label{metrict}
 g^\prime_{RS}(\hat{y}) \frac{\partial \hat{y}^R}{\partial \hat{x}^M}
 \frac{\partial \hat{y}^S}{\partial \hat{x}^N} = g_{MN}(\hat{x}).
\end{equation}
Let us consider infinitesimal coordinate transformations
\[
\hat{y}^M (x) = \hat{x}^M + \hat \kappa \xi^M (\hat{x}).
\]
Representing the initial and transformed metrics as
\bea
g_{MN}(\hat{x})& =& \gamma_{MN}(\hat{x}) +
\hat \kappa h_{MN}(\hat{x}), \nonumber \\
g^\prime_{MN}(\hat{y}) &=& \gamma_{MN}(\hat{y}) +
\hat \kappa h^\prime_{MN}(\hat{y}), \nonumber
\end{eqnarray}
substituting these formulas into Eq. (\ref{metrict}),
and keeping the terms up to the first order in $ \hat \kappa$ we
arrive at the following transformation law for the variation $h_{MN}$:
\begin{equation}\label{gaugetr0}
 h^\prime_{MN}(\hat{x}) =  h_{MN}(\hat{x}) -
\left( \nabla_M \xi_N + \nabla_N \xi_M\right)
\end{equation}
(see, for example, Ref. \cite{Gri}).
It is easy to check that the action built with the second
variation Lagrangian (\ref{Lagrangian0}) is invariant under
transformations (\ref{gaugetr0}) which, therefore, can be interpreted
as gauge transformations of the field $h_{MN}$.

Suppose that the background metric $\gamma_{MN}$ is a
solution of the Einstein equations
\[
 {\cal R}_{MN} - \frac{1}{2} \gamma_{MN}
\left({\cal R}-\Lambda\right) = 8 \pi \hat G T_{MN}
\]
with some energy-momentum tensor of the matter $T_{MN}$.
Expressing ${\cal G}_{MN}$ and $\left({\cal R}-\Lambda\right)$
in terms of the energy-momentum tensor  $T_{MN}$, substituting
it into (\ref{Lagrangian0}) and transforming the kinetic term to the
standard Fierz-Pauli form, we obtain the following second variation
Lagrangian
\begin{eqnarray}\label{Lagrangian01}
& & L^{(2)}_g/\sqrt{-\gamma} = -\frac{1}{4}\left(\nabla_R h_{MN}
\nabla^R h^{MN} -
\nabla_R h \nabla^R h + 2\nabla_M h^{MN}\nabla_N h -\right.
 \\ \nonumber
& & - \left. 2 \nabla_M h^{MN} \nabla^R h_{RN}\right)
 + \frac{1}{2}  h^{MN} h^{PQ} {\cal R}_{MPNQ} -
\frac{1}{2}  h^{MN} h_{NP} {\cal R}_{M}^P+
\\ \nonumber
& & +\frac{\Lambda}{2(d+2)}\left(h_{MN} h^{MN} -\frac{1}{2} hh\right) -
\frac{4 \pi \hat G}{d+2} T_R^R \left(h_{MN} h^{MN} -
\frac{1}{2} hh\right)+ \\ \nonumber
& &+ \left(8\pi \hat G T^{MN} h_{MR} h_N^R - 4\pi \hat G T^{MN} h_{MN} h
\right),
\end{eqnarray}
where ${\cal R}_{MNPQ}$ and  ${\cal R}_{M}^P$ are the curvature
and the Ricci tensors of the metric $\gamma_{MN}$ respectively.
In the next section this expression will be used for
calculation of the second variation Lagrangian in
Randall-Sundrum model.

\section{Second variation Lagrangian in the Randall - Sundrum model}

We write the perturbed metric around the RS solution $\gamma_{MN}$,
Eq. (\ref{metricrs}), in the form given by Eq. (\ref{metricpar})
with $\hat\kappa = M^{-3/2}$, where, as in the Introduction,
$M$ is the fundamental mass scale ("Planck mass") of the
five-dimensional gravity.

In the RS1 model the brane action terms, Eqs. (\ref{RSb1}), (\ref{RSb2}),
give rise to the energy-momentum tensor equal to
\[
 T_{MN} = -\frac{3k}{4\pi \hat G} \sqrt{\frac{\tilde \gamma}{\gamma}}
 \gamma_{\mu\nu} \delta^\mu_M \delta^\nu_N
 \left[\delta (y) - \delta (y-R) \right],
\]
where $\tilde{\gamma}=\det (\gamma_{\mu \nu})$.
The contribution of the gravitational term $S_{g}$, Eq. (\ref{RSg}),
is given by formula (\ref{Lagrangian01}), whereas
the contribution of the branes is equal to
\[
 \Delta L^{(2)}_{1,2} = 3k \left(h_{\mu\nu} h^{\mu\nu} -\frac{1}{2}
\tilde{h} \tilde{h} \right) \sqrt{-\tilde{\gamma}}
\left[\delta (y) - \delta (y-R) \right],
\]
where $\tilde h = \gamma^{\mu\nu} h_{\mu\nu}$. The complete
quadratic Lagrangian for the  variation $h_{MN}$ can be
written as follows:
\begin{eqnarray}
& & L/\sqrt{-\gamma} =  -\frac{1}{4}\left(\nabla_R h_{MN} \nabla^R h^{MN} -
\nabla_R h \nabla^R h + 2\nabla_M h^{MN}\nabla_N h -\right.
 \nonumber \\
& & - \left. 2 \nabla_{M} h^{MN} \nabla^{R} h_{RN}\right)
+\frac{k^2}{2}( h_{MN} h^{MN}+ hh) +  \left[ -2k h_{MN} h^{MN} + \right.
\label{Lagrangian1} \\
& & + \left. k h\tilde h -
k h_{M\nu} h^{M\nu} + 3k( h_{\mu\nu} h^{\mu\nu} -
\frac{1}{2} \tilde h \tilde h)\right]
\left( \delta (y) - \delta (y-R) \right).  \nonumber
\end{eqnarray}
As in Sect. 2, the covariant derivatives $\nabla_{M}$ are calculated
with respect to the background metric $\gamma_{MN}$ of the RS solution
and ${h} = \gamma^{MN} h_{MN}$.

This Lagrangian is invariant under gauge transformations
(\ref{gaugetr0}). In the case under consideration they can be
found explicitly and turn out to be
\begin{eqnarray}\label{gaugetrRS}
h'_{\mu\nu}\left(x,y\right)&=&h_{\mu\nu}\left(x,y\right)-\left(
\partial_{\mu}\xi_{\nu} +\partial_{\nu}\xi_{\mu}+
2\gamma_{\mu\nu}\partial_{4}\sigma\xi_{4} \right)\\ \nonumber
h'_{\mu4}\left(x,y\right)&=&h_{\mu4}\left(x,y\right)-\left(
\partial_{\mu}\xi_{4} +\partial_{4}\xi_{\mu}-2\partial_{4}\sigma\xi_{\mu}
\right)\\ \nonumber
h'_{44}\left(x,y\right)&=&h_{44}\left(x,y\right)-2\partial_{4}\xi_{4},
\end{eqnarray}
where the functions $\xi^M(x,y)$ satisfy the orbifold symmetry conditions
\begin{eqnarray}\label{orbifoldsym1}
\xi^{\mu}\left(x,-y\right)&=&\xi^{\mu}\left(x,y\right)\\
\nonumber
\xi^{4}\left(x,-y\right)&=&-\xi^{4}\left(x,y\right).
\nonumber
\end{eqnarray}
These gauge transformations in other parameterizations were discussed in
papers \cite{CGR} and \cite{AIMVV}.
We will use them to remove the gauge degrees of
freedom of the field $h_{MN}$. To this end we first make a gauge
transformation with
\[
\xi_{4}\left(x,y\right)=\frac{1}{4}\int_{-y}^{y}
h_{44}\left(x,y'\right)dy' -
\frac{x^{4}}{4 R}\int_{-R}^{R}h_{44}\left(x,y' \right)dy' .
\]
One can  easily see that $\xi_{4}$ satisfies the orbifold symmetry
condition. After this transformation $h_{44}$ takes the form
\[
h'_{44}\left(x\right)=\frac{1}{2R}\int_{-R}^{R}
h_{44}\left(x,y' \right)dy'
\]
and therefore it does not depend on $y$. Moreover, there are
no residual gauge transformations involving $\xi_{4}$.

Now let us consider  the components $h_{\mu 4}$. Due to the
orbifold symmetry
\[
h_{\mu4}\left(x,-y\right)=-h_{\mu4}\left(x,y\right),
\]
the gauge transformations for $h_{\mu4}$ read
\[
h'_{\mu4}\left(x,y\right)=h_{\mu4}\left(x,y\right)-\left(
\partial_{4}\xi_{\mu}-2\partial_{4}\sigma\xi_{\mu}\right).
\]
The condition $h_{\mu 4}=0$ is imposed by performing gauge transformations
(\ref{gaugetrRS}) with $\xi_{4}=0$ and $\xi_{\mu}$ satisfying
\[
\xi_{\mu}\left(x,y\right)=e^{2\sigma(y)}\int_{0}^{y}e^{-2\sigma(y')}
h_{\mu4}\left(x,y'\right)dy'.
\]
Note that it satisfies orbifold symmetry condition (\ref{orbifoldsym1}).
Thus we have imposed the gauge
\[
h_{\mu4} =0, \, h_{44} = \phi (x^\mu),
\]
i.e. $h_{44}$ is a function of four-dimensional coordinates
only. We will call it { \it the unitary gauge}. There still remain gauge
transformations satisfying
\begin{equation}\label{remgaugetr}
\partial_{4}\left(e^{-2\sigma}\xi_{\mu}\right)=0.
\end{equation}
They will be important for removing the gauge degrees of freedom of
the massless mode of the gravitational field.

Varying Lagrangian (\ref{Lagrangian1}) we get the equations of motion
for the fluctuations $ h_{MN}(x)$ which, in the
unitary  gauge, reduce to  the following equations for
the fields $h_{\mu \nu}(x,y)$ and $\phi (x)$:
\bea
& &\frac{1}{2}\left(\partial_\rho \partial^\rho h_{\mu\nu}-
\partial_\mu \partial^\rho
h_{\rho\nu}-\partial_\nu \partial^\rho h_{\rho\mu} +
h_{\mu \nu}'' \right)- 2k^2  h_{\mu\nu} +
\frac{1}{2}\partial_\mu \partial_\nu \tilde h +
\frac{1}{2}\partial_\mu \partial_\nu \phi   \nonumber \\
& & + \frac{1}{2} \gamma_{\mu\nu}
\left(\partial^\rho \partial^\sigma h_{\rho\sigma} -
\partial_\rho \partial^\rho \tilde h -  \tilde{h}'' -
4 \sigma' \tilde{h}'
 - \partial_\rho \partial^\rho \phi + 12 k^2 \phi\right)  \nonumber \\
& & + \left[2k  h_{\mu\nu} - 3k\gamma_{\mu\nu}\phi \right]
\left( \delta (y) - \delta (y-R) \right) = 0,   \label{mu-nu}  \\
& & (\partial_\mu \tilde{h} - \partial^\nu  h_{\mu\nu})'-
3 \sigma' \partial_\mu \phi = 0,    \label{mu-4} \\
& & \frac{1}{2}(\partial^\mu \partial^\nu  h_{\mu\nu} - \partial_\mu
\partial^\mu \tilde h ) - \frac{3}{2} \sigma' \tilde{h}'
+ 6 k^2 \phi =0,    \label{4-4}
\eea
where the prime denotes the derivative w.r.t. $y$. In the chosen gauge
Eq. (\ref{mu-4}) is a constraint.
Multiplying Eq. (\ref{4-4}) by 2 and subtracting it from Eq. (\ref{mu-nu})
contracted with $\gamma^{\mu \nu}$ we obtain an auxiliary equation
\begin{equation}\label{contracted-44}
\tilde{h}'' + 2 \sigma' \tilde{h}'- 8k^2 \phi +
\partial_\mu \partial^\mu \phi +
8k \phi \left[ \delta (y) - \delta (y-R) \right] = 0,
\end{equation}
which will be used later.

When written in terms of $h_{\mu \nu}(x,y)$ and $\phi (x)$
the second variation Lagrangian (\ref{Lagrangian1})
is not diagonal, correspondingly the equations of motion,
Eqs. (\ref{mu-nu}) - (\ref{4-4}), are coupled.
To diagonalize the Lagrangian and to decouple the equations we
write the multidimensional gravitational field as
\bea
 h_{\mu\nu}(x,y) & = & b_{\mu\nu}(x,y) + \gamma_{\mu\nu}(y)
 (\sigma (y) - c)\phi (x) \nonumber \\
 & + & \frac{1}{2k^2} \left[ \sigma (y) - c +\frac{1}{2} \right]
 \partial_\mu \partial_\nu \phi (x) +
 \frac{c}{4k^{2}} e^{-2\sigma (y)} \partial_\mu \partial_\nu \phi (x),
 \label{substitution}
\eea
where $c$ is a constant which will be fixed shortly. We will see that
the field $b_{\mu\nu}(x,y)$ describes the massless graviton \cite{RS1,RS2}
and massive Kaluza-Klein spin-2 fields, whereas $\phi (x)$ describes a
scalar field called  the radion.
Apparently, the radion as a massless particle was first identified in
Ref.~\cite{AHDMR01} (see also \cite{Sund98}) and studied in
articles~\cite{radion,CGR}. Actually, substitution
(\ref{substitution}) is suggested by the form of the gauge
transformation which transforms the theory from the local Gaussian normal
coordinates (i.e. coordinates corresponding to $g_{44}=1$, $g_{\mu4}=0$
\cite{CGR}), for which the second variation Lagrangian is diagonal in the
bulk by construction, to our coordinates. Evidently, just a
gauge transformation cannot diagonalize the Lagrangian, because it is gauge
invariant. To carry out the diagonalization and to decouple the equations the
last term in Eq. (\ref{substitution}), the one $\propto e^{-2\sigma}$,
is needed.

It is easy to check that even with arbitrary $c$ Eqs. (\ref{mu-nu}) -
(\ref{4-4}) decouple in the bulk, i.e. for $0 < y < R$.
The junction conditions, coming from Eq. (\ref{mu-nu}), are
\bea
& & b'_{\mu \nu}(x,y_{i}) - \frac{c}{2k^{2}} \sigma' (y_{i})
e^{-2\sigma (y_{i})} \partial_{\mu} \partial_{\nu} \phi (x)  \nonumber \\
& & + \epsilon_{i} \left[ 2k b_{\mu \nu} (x,y_{i}) + \left(
\frac{c}{2k} e^{-2\sigma (y_{i})} + \frac{\sigma (y_{i}) - c}{k} \right)
\partial_{\mu} \partial_{\nu} \phi (x) \right] = 0,  \nonumber
\eea
where $i=1,2$, $y_{1}=0$, $y_{2}=R$, $\epsilon_{1}=+1$, $\epsilon_{2}=-1$.
At $y=0$ the terms $\propto \partial_{\mu} \partial_{\nu} \phi $ vanish
for arbitrary $c$. To make them vanish at $y=R$ we choose
\begin{equation}
  c= \frac{kR}{e^{2kR} -1}.   \label{constant}
\end{equation}
Substituting expression (\ref{substitution}) into
Eq. (\ref{contracted-44}) we get the equality
\[
 \left( e^{2\sigma} \tilde{b}' \right)' = - 2 c
\partial_\mu \partial^\mu \phi.
\]
Making the Fourier expansion in $y$ on the orbifold and taking
into account that the l.h.s. does not contain the zero mode,
we arrive at the conclusion that the radion field satisfies the equation
\beq
\partial_\mu \partial^\mu \phi = 0   \label{radioneq}
\eeq
and $e^{2\sigma}(e^{-2\sigma}b)'$ is a function of $x$ only.
Applying the same argument the latter implies that $(e^{-2\sigma}b)'=0$
(see details in Ref. \cite{BKSV}). By performing transformations
which satisfy (\ref{remgaugetr}) one can impose the gauge
\beq
\tilde{b} = b =0.   \label{T}
\eeq
It is easy to see that there still remain gauge transformations
parameterized by
$\xi_\mu = e^{2\sigma}\zeta_\mu(x)$ with $\zeta_\mu(x)$
satisfying $\partial^\mu \zeta_\mu = 0$.
Substituting expression (\ref{substitution}) into Eqs.
(\ref{mu-4}), (\ref{4-4}) and passing to gauge (\ref{T}),
we arrive at the following relations:
$\partial^\mu \partial^\nu  b_{\mu\nu}=0$ and
$(e^{-2\sigma} \partial^\mu  b_{\mu\nu})'=0$.
The remaining gauge transformations are sufficient to impose the
condition
\begin{equation}
 \partial^\mu  b_{\mu\nu} = 0.   \label{gaugecond}
\end{equation}
Conditions (\ref{T}) and (\ref{gaugecond}) define
the gauge often called the transverse-traceless (TT) gauge.
Having imposed this gauge, we are still left with residual
gauge transformations
\begin{equation}
\label{gaugetr}
\xi_\mu = e^{2\sigma}\zeta_\mu(x), \quad
\partial_\rho \partial^\rho \zeta_\mu = 0,
\quad \partial^\mu \zeta_\mu = 0,
\end{equation}
which, as we will see shortly, are important for
determining the number of degrees of freedom of the
massless mode of $b_{\mu\nu}$.

In this gauge Eqs.  (\ref{mu-4}), (\ref{4-4}) are trivially
satisfied. Eq. (\ref{mu-nu}) becomes
\[
  \frac{1}{2} \partial_\rho \partial^\rho b_{\mu\nu} +
  \frac{1}{2} b_{\mu\nu}''  -
  2k^2 b_{\mu\nu} +  2k b_{\mu\nu} \left( \delta (y) -
  \delta (y-R) \right) = 0.
\]
Thus, we have decoupled the equations of motion.

Now let us consider the second variation action of the theory.
Substituting (\ref{substitution}) with $c$ given by
(\ref{constant}) into Lagrangian (\ref{Lagrangian1}) and  taking into
account the TT gauge conditions (\ref{T}),
(\ref{gaugecond}) for $b_{\mu\nu}$, we get
\bea
  L/\sqrt{-\gamma} & = & -\frac{1}{4} \partial_\rho  b_{\mu\nu} \partial^\rho
  b^{\mu\nu}  -\frac{1}{4}( \partial_4  b_{\mu\nu} -2\partial_4 \sigma
  b_{\mu\nu})( \partial_4  b^{\mu\nu} +2\partial_4 \sigma
  b^{\mu\nu})   \nonumber \\
  & - & \frac{3}{4}  c\,
  e^{-2 \sigma}\partial_\mu \phi\partial^\mu \phi.
             \label{Lagrangian3}
\eea
Integrating over $y$ we obtain the
following four-dimensional effective Lagrangian for the field $\phi$:
\[
L_{\phi} = - \frac{3}{4}c \int_{-R}^{R} e^{-2\sigma (y)}
\sqrt{-\gamma} \partial_{\mu} \phi \partial^{\mu} \phi \, dy =
-\frac{3}{2}\frac{kR^2}{e^{2kR}-1} \eta^{\mu \nu}
\partial_\mu \phi \partial^\nu \phi.
\]
To bring the kinetic term to the canonical form we rescale the
field according to
\begin{equation}
 \phi (x) = \sqrt{\frac{e^{2kR}-1}{3 kR^2}} \varphi (x).
\label{rescaling-phi}
\end{equation}

It remains to decompose the field $b_{\mu\nu}$ into Kaluza-Klein
modes with definite masses and to integrate Eq. (\ref{Lagrangian3})
over $y$.
Following Refs. \cite{RS2,DHR1,DHR2} we write
\begin{equation}
b_{\mu\nu}(x,y) = \sum_{n=0}^{\infty}
b^{(n)}_{\mu\nu}(x) \psi_n(y),   \label{decomp}
\end{equation}
where $\{ \psi_{n}(y), \; \; n=0,1,2,\ldots\}$ form the complete
orthogonal system of eigenfunctions of the equation
\[
 \left[\frac{1}{2} \frac{d^2}{dy^2} +
 2k(\delta(y) - \delta(y-R)) - 2k^2 \right] \psi_{n}(y) = -
 \frac{m_{n}^{2}}{2} e^{2k|y|} \psi_{n}(y)
\]
satisfying the normalization condition
\[
\int_{-R}^R e^{2k|y|} \psi_m (y) \psi_n (y)dy = \delta_{mn}.
\]
The solutions to the eigenvalue problem are equal to
\beq
\psi_0 (y) = N_{0}\, e^{-2k|y|}, \; \; \;
\psi_n (y) = N_n \Psi_{n}(y) \; \; (n=1,2, \ldots), \label{zeromass}
\eeq
where
\[
\Psi_{n}(y) = C_{1}J_2 \left(\beta_{n}e^{-(\sigma(y)+kR)}\right)
  + C_{2} Y_2 \left( \beta_{n}e^{-(\sigma(y)+kR)} \right)
\]
and $J_2(\xi)$ and $Y_2(\xi)$ are the Bessel and Neumann functions
respectively. The coefficients $C_{1}$, $C_{2}$ and the values of
$\beta_{n}$ are determined by the junction conditions
at $y=0$ and $y=R$. One gets that
\[
C_{1} = Y_1 \left(\beta_{n}e^{-kR} \right), \; \; \;
C_{2} = - J_1 \left(\beta_{n}e^{-kR}\right)
\]
and $\beta_{n}$ are roots of the equation
\[
J_1(\beta_{n} e^{-kR})Y_1(\beta_{n}) -
J_1(\beta_{n}) Y_1(\beta_{n} e^{-kR}) =0.
\]
As we will see shortly, $e^{-kR} \sim 10^{-15}$ so that for small
$n$ the numbers $\beta_{n}$ satisfy approximately the equation
$J_{1}(\beta_{n}) = 0$. The values of the first few $\beta_{n}$
are given by $\beta_{n}=3.83,7.02,10.17,13.32$
for $n=1,2,3,4$ respectively.
The normalization constants in Eq. (\ref{zeromass}) are found to be
\beq
 N_{0} = \frac{ \sqrt{k}}{\sqrt{1-e^{-2kR}}},
 \; \; \; N_{n} = \frac{ \sqrt{k}}{\sqrt{e^{2kR}\Psi_n^2(\beta_{n})-
 \Psi_n^2(\beta_{n} e^{-kR})}}.   \label{N0Nn}
\eeq
For small $n$ the values of functions (\ref{zeromass})
at $y=0$ and $y=R$ can be estimated as
\bea
  \psi_{0} (0) & = & N_{0} \approx \sqrt{k}; \; \; \;
  \psi_{n} (0) \approx \frac{\sqrt{k}}{|J_{2}(\beta_{n})|} e^{-kR},
  \; \; (n \neq 0) \label{psi-0} \\
 \psi_{0} (R) & = & N_{0} e^{-2kR} \approx \sqrt{k} e^{-2kR}; \; \; \;
  \psi_{n} (R) \approx - \sqrt{k} e^{-kR} \; \; (n \neq 0). \label{psi-1}
\eea

Performing mode expansion (\ref{decomp}) we obtain the
following expression
for the four-di\-men\-si\-o\-nal effective action
({\it i.e.} dimensionally reduced action) of the five-dimensional
gravitational field in the RS model:
\bea
S_{eff} & = & \int \left[ -\frac{1}{4}\sum_{n=0}^{\infty}
\left(\partial_{\mu}b^{(n)}_{\rho\sigma}
\partial_{\nu}b^{(n)}_{\epsilon \delta} \eta^{\mu \nu}
\eta^{\rho \epsilon} \eta^{\sigma \delta} +
{m_n^{2}} b^{(n)}_{\mu \rho}b^{(n)}_{\nu\sigma}
\eta^{\mu \nu} \eta^{\rho \sigma} \right)  \right. \nonumber \\
 & - & \left. \frac{1}{2} \eta^{\mu \nu}
\partial_{\mu}\varphi \partial_{\nu}\varphi\right] dx, \label{effaction}
\eea
where $m_{n} = \beta_{n} k e^{-kR}$.
We see that the effective theory includes a massless spin-2 tensor field
(the graviton), an infinite tower of massive
spin-2 tensor  fields (often referred to as massive KK gravitons) and
a massless scalar field (the radion) whose classical equation
of motion is Eq. (\ref{radioneq}). An important
point is that, due to  the form of the zero mode function
(see Eq. (\ref{zeromass})), remaining gauge freedom (\ref{gaugetr})
allows transformations
\[
  b^{(0)}_{\mu\nu}(x) \rightarrow
  b^{\prime 0}_{\mu\nu}(x)=  b^0_{\mu\nu}(x) -
  (\partial_\mu \zeta_\nu(x)+\partial_\nu
  \zeta_\mu(x))
\]
of the field $b^{(0)}_{\mu\nu}(x)$ with $\zeta_{\mu} (x)$ satisfying
$\partial^\mu \zeta_\mu(x) =0$,
$\partial_\mu \partial^\mu \zeta_\nu(x) =0$.
This guarantees that the field $ b^{(0)}_{\mu\nu}(x)$ has
only two degrees of freedom and, therefore, can indeed be
identified with the field of the massless graviton.

It turns out that the five-dimensional gravity, described by
effective action (\ref{effaction}), looks different,
when viewed from different branes. One of the ways to see this property
is to consider the interaction of the five-dimensional gravity
with matter on the branes. This problem will be addressed in the
next section.

\section{Effective theories on the branes}

To begin with let us consider a scalar field $\Phi (x)$ localized on
brane $i$ $(i=1,2)$. Suppose that its dynamics is determined by the
standard action with quartic coupling
\beq
S_{\Phi,i} = -\int_{B_{i}} dx \sqrt{-\gamma (x,y_{i})}
\left[ \frac{1}{2} \gamma^{\mu \nu}(x,y_{i}) \partial_{\mu} \Phi
\partial_{\nu} \Phi +\frac{m^{2}}{2} \Phi^{2} +
\frac{\lambda}{4} \Phi^{4} \right],          \label{SB-1}
\eeq
where, as before, $i=1,2$ and $y_{1}=0$, $y_{2}=R$. Here we consider the
leading approximation in $\hat{\kappa}$, therefore fluctuations
$h_{MN}$ (see Eq. (\ref{metricpar})) are neglected. The induced
metric on brane $i$ is equal to
\bea
  \gamma_{\mu \nu} & = & e^{2\sigma_{i}} \eta_{\mu \nu}, \label{gamma-i} \\
  \sigma_{1} & \equiv & \sigma (0) = 0, \; \; \;
   \sigma_{2} \equiv \sigma (R) = -kR.   \nonumber
\eea

Action (\ref{SB-1}) can be rewritten as
\beq
S_{\Phi,i} = -\int_{B_{i}} dx e^{4\sigma_{i}}
\left[ \frac{1}{2} e^{-2\sigma_{i}} \eta^{\mu \nu}
\partial_{\mu} \Phi \partial_{\nu} \Phi + \frac{m^{2}}{2} \Phi^{2} +
\frac{\lambda}{4} \Phi^{4} \right].          \label{SB-2}
\eeq
The conventional interpretation is obtained by rescaling
the field
\[
\Phi (x) \rightarrow e^{-\sigma_{i}} \Psi (x)
\]
in order to bring the kinetic term to the canonical form. One gets
\[
S_{\Phi,i} = -\int_{B_{i}} dx \left[ \frac{1}{2} \eta^{\mu \nu}
\partial_{\mu} \Psi \partial_{\nu} \Psi +
\frac{1}{2} m^{2}e^{2\sigma_{i}} \Psi^{2} +
\frac{\lambda}{4} \Psi^{4} \right].
\]
The conclusion, derived on the basis of this formula, states
that the mass of the field in the effective theory on brane 2 is
$\tilde{m} = me^{-kR}$. With $m \sim M_{Pl}$ and
$kR \approx 30 \div 35$ the brane mass $\tilde{m} \sim 1$TeV, i.e.
the TeV-scale is obtained from the Planck scale due to the
exponential factor \cite{RS1}.

However, the induced metric on brane 2, Eq. (\ref{gamma-i}), contains
the exponential factor $e^{2\sigma_{2}} = e^{-2kR}$ which is
exactly of the same origin as the one discussed above, namely it also
comes from the warp factor in the background metric, Eq. (\ref{metricrs}).
The coordinates $\{x^{\mu}\}$ therefore are not Galilean on this brane.
As we already mentioned in the Introduction, coordinates
are called Galilean if the metric tensor is $diag(-1,1,1,1)$, see
Ref. \cite{LL}.

In Ref. \cite{GNS} the correct physical mass of a brane field was
determined from the asymptotic behavior of the 2-point function expressed
in terms of the proper distance. Here we show that the correct
interpretation of the effective theory on brane 2 and correct determination
of the mass can be achieved simply by changing to the Galilean
coordinates
\beq
  x^{\mu} \rightarrow z^{\mu} = e^{-kR} x^{\mu}.   \label{change}
\eeq
In these coordinates the action $S_{\Phi,2}$, Eq. (\ref{SB-2}), is equal to
\beq
S_{\Phi,2} = -\int_{B_{2}} dz
\left[ \frac{1}{2} \eta^{\mu \nu}
\frac{\partial}{\partial z^{\mu}} \tilde{\Phi}
\frac{\partial}{\partial z^{\nu}} \tilde{\Phi}
+ \frac{m^{2}}{2} \tilde{\Phi}^{2} +
\frac{\lambda}{4} \tilde{\Phi}^{4} \right],          \label{SB-4}
\eeq
where $\tilde{\Phi}(z) = \Phi (x)$. Therefore, the brane field on
brane 2 has the physical mass equal to its Lagrangian mass $m$. In fact,
expression (\ref{SB-4}) is very easy to understand. Indeed, since
Eq. (\ref{SB-1}) is covariant, by changing the coordinates $\{x^{\mu}\}$
to $\{z^{\mu}\}$ in it we arrive immediately at Eq. (\ref{SB-4}).

The general form of the interaction of the five-dimensional
gravity with matter is given by the following standard expression:
\begin{equation}
\frac{ \hat \kappa}{2} \int_{B_1} h^{\mu\nu}(x,0) T^{(1)}_{\mu\nu} dx +
\frac{ \hat \kappa}{2} \int_{B_2} h^{\mu\nu}(x,R) T^{(2)}_{\mu\nu}
 \sqrt{- \det \gamma_{\mu \nu}(R)} dx,     \label{interaction}
\end{equation}
where $ T^{(1)}_{\mu\nu}$ and $ T^{(2)}_{\mu\nu}$ are the
energy-momentum tensors of the matter on brane 1 and brane 2
respectively. Substituting (\ref{substitution}) into
(\ref{interaction}), decomposing $b_{\mu\nu}(x,y)$ according to
(\ref{decomp}) and rescaling the field $\phi$
as in Eq. (\ref{rescaling-phi}), we find that the interaction on
brane 1 is given by
\begin{equation}
\frac{1}{2} \int_{B_1} \left[ \kappa_1   \, b^{(0)}_{\mu\nu}(x)
 T^{(1)\mu\nu}+   \kappa_2 \sum_{n =1}^\infty \omega_{n}^{(1)}
 b^{(n)}_{\mu\nu}(x)
 T^{(1)\mu\nu}-\frac{\kappa_{2}}{ \sqrt{3}}
 \varphi\, T^{(1)\mu}_{\mu} \right] dx.
\label{intb}
\end{equation}
The coupling constants of the zero mode, the radion and the massive KK modes
on brane 1 are equal to $\kappa_{1}$, $\kappa_{2} \omega^{(1)}_{n}$ and
$(-\kappa_{2}/\sqrt{3})$ respectively, where
\beq
\kappa_{1} = \hat{\kappa} N_0 \sim \frac{\sqrt{k}}{M^{3/2}},
\; \; \;  \kappa_{2} = \kappa_{1} e^{-kR},   \; \; \;
\omega_{n}^{(1)} = \frac{\psi_{n}(0)}{N_{0}} e^{kR}. \label{coupl1}
\eeq
Here we used relations (\ref{psi-0}), which also imply that
$\omega_{n}^{(1)} \sim 1$ for small $n$. In particular,
 $\omega^{(1)}_{n} = 2.48, 3.34, 4.00, 4.58 \ldots$ for $n=1,2,3,4, \ldots$.

Now let us analyze the interaction of the five-dimensional gravity
with matter on brane 2. Again,
to have the correct interpretation of the effective theory on brane 2
we change the coordinates $\{x^{\mu}\}$ for the Galilean
coordinates $\{z^{\mu}\}$, Eq. (\ref{change}).
In these coordinates the relevant term in Eq. (\ref{interaction})
takes the form
\beq
\frac{ \hat \kappa}{2} \int_{B_2} h^{\prime}_{\mu\nu}(z,R)
T^{\prime (2)}_{\rho \sigma} \eta^{\mu \rho} \eta^{\nu \sigma} dz,
         \label{intb-prime}
\eeq
where $T^{\prime (2),\mu\nu}$ is the canonical energy-momentum tensor of
the matter,
\[
h^{\prime}_{\mu\nu}(z,R) = e^{2kR} h_{\mu\nu}(x,R),
\]
and the prime labels tensors in the coordinates $\{ z^{\mu} \}$.
{}From now on the four-dimensional indices will be raised and
lowered using the Minkowski metric tensor $\eta_{\mu \nu}$.

To get the correct couplings we interprete action
(\ref{effaction}) as the effective action of the
bulk graviton fields and radion on brane 2. In Galilean
coordinates (\ref{change}) it becomes
\bea
S_{eff} & = & \int \left[ -\frac{1}{4}\sum_{n=0}^{\infty}
\left( e^{-2kR}
\frac{\partial}{\partial z^{\mu}} b^{\prime (n)}_{\rho\sigma}
\frac{\partial}{\partial z_{\mu}} b^{\prime (n),\rho\sigma}+
{m_n^{2}} b^{\prime (n)}_{\rho\sigma}b^{\prime (n),\rho\sigma}\right)
\right. \nonumber \\
& - & \left. \frac{1}{2} e^{2kR} \frac{\partial}{\partial z^{\mu}}
\varphi^\prime
\frac{\partial}{\partial z_{\mu}} \varphi^\prime \right] dz.  \nonumber
\eea
Rescaling the graviton field and the radion field as
\beq
b^{\prime (n)}_{\rho\sigma} (z)= e^{kR} u^{(n)}_{\rho\sigma}(z), \; \; \;
\varphi^\prime (z)= e^{-kR} \chi (z)    \label{rescale-bphi}
\eeq
respectively, we obtain the effective action in the canonical form
\bea
S_{eff} & = & \int \left[ -\frac{1}{4} \sum_{n=0}^{\infty}
\left( \frac{\partial}{\partial z^{\mu}} u^{(n)}_{\rho\sigma}
\frac{\partial}{\partial z_{\mu}} u^{(n) \rho\sigma}+
{(m_n e^{kR})^{2}} u^{(n)}_{\rho\sigma} u^{(n) \rho\sigma} \right)
   \right.  \nonumber \\
& - & \left. \frac{1}{2} \frac{\partial}{\partial z^{\mu}} \chi
\frac{\partial}{\partial z_{\mu}} \chi \right] dz.
\label{effaction2}
\eea
We see that the masses of the gravitons as measured by an observer on
brane 2 are equal to
\beq
     M_{n} = m_{n} e^{kR} = k \beta_{n}.   \label{Mn}
\eeq
Contrary to the case of brane fields described by action (\ref{SB-1}),
the graviton mass spectrum is multiplied by the scale factor $e^{kR}$
when we pass from the coordinates $\{x^{\mu}\}$ to $\{z^{\mu}\}$. The
reason is that expression (\ref{effaction}) is not covariant, namely the
indices are raised with the Minkowski metric whereas the metric
in the coordinates $\{x^{\mu}\}$ is
$\gamma_{\mu \nu} (x,R) = e^{-2kR} \eta_{\mu \nu}$.

We would like to note that in fact relation (\ref{Mn}) between the
Lagrangian mass and the mass measured by an observer on brane 2
is valid for bulk fields of any tensor structure (see an
example in \cite{GNS}).

Using Eq. (\ref{substitution}) and making the mode decomposition
accompanied by rescaling (\ref{rescale-bphi}) interaction term
(\ref{intb-prime}) on brane 2 takes the form
\bea
& & \frac{1}{2} \int_{B_2} \left[ \hat{\kappa} e^{kR} \psi_{0}(R)
 u^{(0)}_{\mu\nu}(z) T^{(2) \mu\nu} +
 \hat{\kappa} \sum_{n=1}^{\infty} e^{kR} \psi_{n}(R) u^{(n)}_{\mu\nu}(z)
 T^{(2) \mu\nu} \right. \nonumber \\
& & \left. - \frac{\hat{\kappa}}{\sqrt{3}}
 \frac{\sqrt{k}}{\sqrt{1-e^{-2kR}}} \chi T^{(2) \mu}_{\mu} \right] dz
                   \label{pre-inf2}  \\
& & = \frac{1}{2} \int_{B_2} \left[ \kappa_2 \,  u^{(0)}_{\mu\nu}(z)
 T^{(2) \mu\nu} - \kappa_1 \sum_{n=1}^\infty \omega_{n}^{(2)}
 u^{(n)}_{\mu\nu}(z) T^{(2) \mu\nu} \right. \nonumber \\
& & \left. -\frac{\kappa_{1}}{ \sqrt{3}}
\chi \, T^{(2) \mu}_{\mu}\right] dz.
 \label{intf2}
\eea
The couplings of the zero mode, massive KK modes and the radion
are $\kappa_{2}$, $(-\kappa_{1} w^{(2)}_{n})$ and $(-\kappa_{1}/\sqrt{3})$
respectively, where $\kappa_1$ and $\kappa_2$ were defined by Eq.
(\ref{coupl1}) and
\bea
\omega_{n}^{(2)} & = & - \frac{\psi_{n}(R) e^{kR}}{N_{0}}. \nonumber
\eea
Using Eqs. (\ref{psi-1}) it can be easily checked that
$\omega_{n}^{(2)} \approx 1.0$ for small $n$.

We see that the order of magnitude of the graviton and radion
interactions with matter on both branes are set
by two parameters. We have chosen them to be $\kappa_1$ and
$\kappa_2$, the couplings of the massless graviton to mater on brane
1 and brane 2 respectively. They differ in the exponential factor
$e^{-kR}$ (see Eq. (\ref{coupl1})).

Let us discuss the effective models on brane 1 and brane 2.
In the conventional and physically interesting scenario it is assumed
that our brane is brane 2, whereas brane 1 accommodates a mirror world.
To have usual Newton's gravity mediated by the massless graviton we
identify its coupling constant $\kappa_{2}$ with $1/M_{Pl}$. Combining
Eqs. (\ref{N0Nn}) and (\ref{coupl1}) we obtain the following relation
between the Planck mass and the fundamental mass scale $M$ (recall that
$\hat{\kappa}^{2} = M^{-3}$):
\begin{equation}\label{cb2}
M_{Pl}^{2} = \frac{M^{3}}{k} \left( e^{2kR}-1 \right) \approx
\frac{M^{3}}{k} e^{2kR}.
\end{equation}
By choosing $M \sim k \sim 1$TeV we reproduce the
correct value of the Planck mass, if the argument of
the exponential factor satisfies $kR \approx 30 \div 35$. From
Eqs. (\ref{coupl1}), (\ref{intf2}) it follows that, as in the
standard scenario, the couplings of the massive KK modes
and of the massless radion to matter on brane 2 are of order
of $1 \; \mbox{TeV}^{-1}$ \cite{DHR1,DHR2}.
The effective Lagrangian (\ref{intf2}) can be rewritten in the following
form, from which the above features are clearly seen:
\[
\frac{1}{2} \int_{B_2}
\left[ \frac{1}{M_{Pl}} u^{(0)}_{\mu\nu}(z) T^{(2) \mu\nu} +
 \sum_{n=1}^\infty
 \frac{\omega_{n}^{(2)}}{\Lambda_{\pi}} u^{(n)}_{\mu\nu}(z)
 T^{(2) \mu\nu}
 - \frac{1}{\Lambda_{\pi} \sqrt{3}}  \chi T^{(2) \mu}_{\mu}\right] dz,
\]
where ${\Lambda_\pi} = M_{Pl}e^{-kR}= \sqrt{M^{3}/k} \sim 1$TeV
and $M_{Pl}$ is given by (\ref{cb2}).
The masses of the KK excitations are given by Eq. (\ref{Mn}). For small
$n$ the masses $M_{n} \sim k \sim 1 \; \mbox{TeV}$.
The presence of the massless radion with such coupling
leads to some predictions, which
are in contradiction with the available high energy physics data.
To avoid this problem several mechanisms for generating a mass for the
radion  were proposed (see, for example, Ref. \cite{wise}).

Now we turn to the analysis of the positive tension mirror brane (brane 1).
The coupling $\kappa_1$ of the massless graviton
is related to the fundamental mass scale $M$ and the
parameter $k$ as
\[
  \kappa_{1}^{-2} = \frac{M^{3}}{k} \left( 1-e^{-2kR} \right).
\]
With the parameters defined by physical conditions imposed on brane 2
the coupling of the massless graviton
$\kappa_{1} \sim 1/\Lambda_{\pi}$, whereas the couplings of the
massive gravitons and of the radion to matter turn out to be
$\sim 1/M_{Pl}$.

Suppose for a moment that brane 1 is our brane.
In this case we would have to require that
$\kappa_{1} \sim M_{Pl}^{-1}$. This can be achieved by
taking $M \sim k \sim M_{Pl}$ and keeping $\exp (kR) \gg 1$.
Then, from relations (\ref{coupl1}), we would get that
$\kappa_1 \gg \kappa_{2}$, i.e.
the interactions of the massive KK modes
and of the radion field with matter on brane 1 would be much weaker
(exponentially suppressed) than that of the massless graviton. Therefore,
on brane 1 the radion field would not affect Newton's gravity, which is
determined by the interaction of the massless graviton in (\ref{intb}).
The massive KK modes have masses of the order of $M_{Pl} e^{-kR}$.
Thus, such scenario would be
phenomenologically acceptable, but not interesting.

\section{Conclusions and discussion}

We presented a consistent Lagrangian description of linearized
gravity in the Randall-Sundrum model with two branes paying
attention to some issues which were skipped or received little
attention in previous publications. In particular,
we discussed the gauge freedom and gauge fixing, and carried
out the diagonalization of the Lagrangian and decoupling
of the classical equations of motion in detail. We also
revised the mode expansion of the bulk gravitational field.
These enabled us to identify the physical degrees of freedom in the
RS1 model and construct the effective Lagrangian. We would like to note
that in the limit $R\rightarrow \infty$ the radion field $\phi$ drops
from the Lagrangian, and we get the same degrees of freedom, as found
in \cite{AIMVV} for the case of infinite extra dimension.

An important point in our studies is that the five-dimensional gravity
looks different when viewed from different branes. This point was stressed
and worked out in some papers, see, for example, \cite{Rub01,GNS}.
Our central observation is that in order to have
the correct interpretation of the effective theories on the branes
and to calculate the values of masses and coupling constants measured by
a brane observer one has to change to the proper (Galilean) coordinates.
We showed that contrary to the conventional interpretation the measured
masses of brane fields coincide with the Lagrangian mass
(see Eq. (\ref{SB-4})). The masses of bulk fields measured on the
brane at $y=R$ acquire the enhancing exponential factor comparing
to the Lagrangian masses, see Eq. (\ref{Mn}). As we have mentioned
in the Introduction, in Ref. \cite{GNS} these mass relations were
obtained from the analysis of the 2-point Green functions. We also
derived the relation between the Planck mass and the fundamental scale
of the five-dimensional gravity, Eq. (\ref{cb2}), which differs
from the conventional one.

As a matter of fact, instead of choosing the Ga\-li\-li\-ean
coordinates for description of the effective theory on brane 2 only, one can
introduce global five-dimensional coordinates $\{ z^{\mu},y \}$, for which
the warp factor  on brane 2 is equal to 1, i.e. $\{z^\mu\}$ are Galilean
coordinates on brane 2.  This alternative description was used in Ref.
\cite{Rub01} for derivation of  relation (\ref{cb2}). All conclusions
concerning the physical content of theories on brane 1 and brane 2,
obtained there, coincide with ours.

Our expression for the Lagrangian of the
effective four-dimensional theory on brane 2 in the physical (Galilean)
coordinates, given by (\ref{effaction2}), (\ref{intf2}), differs from the
one usually discussed in the literature (see, for example, Refs.
\cite{DHR1,DHR2}). In particular, we get $M \sim k \sim 1$TeV, whereas in the
above mentioned  papers one needs $M \sim k \sim M_{Pl}$.
In our case, according to Eq. (\ref{Mn})
the spacings between masses in the spectrum
are $\sim k \sim R^{-1}$, similar to the usual Kaluza-Klein
theories with direct product topology of the space-time. If the
analysis of the RS model is carried out in non-Galilean coordinates,
as it is done in many papers,
then the spacings are $\sim k e^{-kR} \sim R^{-1} e^{-kR} \ll R^{-1}$,
which is valid only on brane 1.

Essentially, the hierarchy, i.e. the existence of two scales separated
by many orders of magnitude,  is generated in the same way as in the
conventional interpretation. Namely, according to Eq. (\ref{cb2})
the scale of five-dimensional theory turns out to be exponentially
suppressed with respect to the scale of four-dimensional gravity on brane 2.
In fact this possibility was mentioned briefly in
Ref. \cite{RS2} (see also \cite{Rub01}). As it can be seen from the
first term in Eq. (\ref{pre-inf2}), the exponential factor
comes from $e^{kR}\psi_{0}(R) \sim \sqrt{k} e^{-kR}$, i.e.
from the exponential decay of the graviton wave function
which is proportional to the warp factor in the background metric.

The relative strengths of interaction of massless and massive gravitons
with matter are explained by different rates of the falloff of the
wavefunctions of these fields along the fifth direction. Indeed, from
Eqs. (\ref{pre-inf2}), (\ref{intf2}) it follows that
\[
  \frac{\kappa_{2}}{\kappa_{1} \omega_{n}} = -
  \frac{\psi_{0}(R)}{\psi_{n}(R)} \sim e^{-kR}.
\]
One last remark. Though formulas (\ref{coupl1}) and (\ref{cb2}),
expressing the couplings $\kappa_{1}$ and
$\kappa_{2}$ in terms of the parameters $M$, $k$ and $R$ of the model,
are different from the ones used in the literature
(see \cite{RS1, DHR1, DHR2}), it is easy to check that the
ratios $\kappa_{1}/\kappa_{2}$ and $\kappa_{1}\omega_n^{(2)}/\kappa_{2}$
are the same both in  physical (Galilean) coordinates $\{ z^{\mu} \}$
and in "non-physical" coordinates $\{ x^{\mu} \}$. For this reason
phenomenological predictions obtained in previous papers remain valid.

\bigskip
{ \large \bf Acknowledgments}
\medskip

The authors are grateful to G.Yu. Bogoslovsky, Yu.V. Grats and V.A. Rubakov
for useful discussions. The work of E.B., Yu.K. and I.V. was supported by
RFBR grant 00-01-00704  and the grant 990588 of the programme "Universities
of Russia". E.B. was supported in part by the CERN-INTAS grant 99-0377 and
by the INTAS grant 01-0679. Yu.K. and I.V. were also supported in part by
the programme SCOPES (Scientific co-operation between Eastern Europe and
Switzerland) of the Swiss National Science Foundation (project No.
7SUPJ062239) and financed by the Swiss Federal Department of Foreign
Affairs.

\end{document}